# Metrology of Complexity and Implications for the Study of the Emergence of Life

Sara Imari Walker[1,2,3*]

[1]School of Earth and Space Exploration, Arizona State University, Tempe AZ USA

[2]Beyond Center for Fundamental Concepts in Science, Arizona State University, Tempe AZ USA

[3]Santa Fe Institute, Santa Fe, NM USA

*author for correspondence: sara.i.walker@asu.edu;

## Abstract

One of the longest standing open problems in science is how life arises from non-living matter. If it is possible to measure this transition in the lab, then it might be possible to understand the physical mechanisms by which the emergence of life occurs, which so far have evaded scientific understanding. A significant hurdle is the lack of standards or a framework for cross comparison across different experimental contexts and planetary environments. In this essay, I review current challenges in experimental approaches to origin of life chemistry, focusing on those associated with quantifying experimental selectivity versus *de novo* generation of molecular complexity, and I highlight new methods using molecular assembly theory to measure molecular complexity. This metrology-centered approach can enable rigorous testing of hypotheses about the cascade of major transitions in molecular order marking the emergence of life, while potentially bridging traditional divides between metabolism-first and genetics-first scenarios. Grounding the study of life's origins in measurable complexity has significant implications for the search for life beyond Earth, suggesting paths toward theory-driven detection of biological complexity in diverse planetary contexts. As the field moves forward, standardized measurements of molecular complexity may help unify currently disparate approaches to understanding how matter transforms to life. Much remains to be done in this exciting frontier.

# Main

The origin of life is one of the most enigmatic open problems across the sciences, with many outstanding questions pertaining to what role physics might play in the transformation from matter to life[1]. Pursuit of a physics of living matter has often emphasized a critical role for information as fundamental[2], with an implication that a focus on informational properties of living matter might help delineate the boundary between non-life and life[3]. Almost a century ago, emphasis on the role of algorithmic information underlying genetic inheritance, stored in an aperiodic crystalline material[4], inspired the search for the structure of DNA[5]. In subsequent decades, as molecular and cell biology advanced, emphasis expanded from a focus on the information content of genes to information processing in networks of interacting molecules[6], the development of entire organisms during morphogenesis[7], in mediating collective behaviors[8], and the process of evolution itself being algorithmic[9]. While ideas on the nature and role of information in life have expanded, the utility of this conceptual framing for making traction on frontier problems, like the emergence of life from non-living matter, remains an open challenge.

Among the most important aspects of physics is metrology, the science of measurement, because it allows standardization and empirical testing of theory. Advances in metrology have driven the incredible predictive power of physics, and historically underlie why physics has unlocked many properties of our universe so successfully[10]. One of the key challenges for origins of life research is that there has been no way to measure – or even define – "life". Early in my career I started to approach the origin of life from an information-first lens, motivated by the idea that the information aspects of life might be key to quantifying the transition from non-living to living matter, such that we might measure the transition from non-living to living matter[11]. However, origin of life science presents unique problems for information-theoretic approaches. These include issues in estimating entropy and information because probabilities cannot be assigned to the likelihood of discovering unobservable molecular configurations or their function, and the challenges presented by post-selecting on extant biochemistry. These challenges render information theoretic quantities not computable in cases of greatest interest to origin of life science, confounding the necessary careful and challenging work required to map theoretical metrics to what we can physically measure in the chemistry lab. It is worth reviewing these challenges, before discussing what advances are possible with newly emerging approaches based on a metrology of complexity.

Fossil evidence suggests life emerged on Earth at least ~3.8 billion years ago[12,13]. Evidence of a cellular architecture consistent with the last universal common ancestor (LUCA) of cellular life is dated at approximately 3.5 billion years ago[14]. LUCA is not believed to have been a single cell but rather represents the collective properties of molecular systems that nucleated the three major lineages of life we now recognize as Archaea, Bacteria and Eukaryota[15]. All major transitions in molecular structure, ranging from the emergence of organized metabolism[16] to homochiral selection of macromolecules[17], to the translation system[18], occurred before LUCA. This suggests an extensive period of molecular evolution preceded the emergence of cellular life as we now understand it.

In prebiotic chemistry, the study of the chemistry that preceded life, a natural target is recreating the chemistry of LUCA from abiotic conditions[19]. Thus, the goal is to navigate missing stages of evolution with a direct bridge linking abiotic chemistry retrosynthetically to the chemistry of LUCA[20]. Experiments often operate within different classes of hypotheses. The historical division has been between metabolism-first[21,22] or genetics-first[23–25] scenarios, roughly corresponding to whether the hypothesis favors selection of a self-reproducing networks composed of small molecules (metabolism), or macromolecular replication (genetics) as the first of the significant early molecular transitions leading to cellular life. Recently there have been calls for more integrated approaches, particularly from early career scientists, who see paths forward that do not fall neatly within these traditional divisions[26,27]. Nonetheless, most prebiotic chemistry experiments target synthesis of coded amino acids[28], ribonucleosides[29], lipids[30], or metabolic precursors[22], depending on the hypothesis tested.

An infrequently discussed, but important challenge for any experimental program on the origin of life is the role of post-selection[31,32] and human intervention[33]. Post-selection is a concept from probability theory[34], where one conditions the probability space on the occurrence of a given event, after the event occurs. In prebiotic chemistry, experiments are almost universally conditioned on the biochemistry of life as we know it[35]. By imposing our expectations from modern biochemistry onto prebiotic chemistry we can dramatically reduce the search space[36], but we do so at significant risk of inputting selective biases that may obscure our ability to infer the probability of life to emerge from unconstrained planetary starting conditions[37]. Selective information input by an experimenter[33,38,39] targeting the synthesis of known biomolecules includes choices made in terms of varying sets of reagents, the experimental conditions, reaction hardware, purification methods, reagent purity, and process parameters such as order of addition, temperature, and residence time. If we are to understand the emergence of life as the emergence of selective circumstances favoring increasing molecular order and complexity, the future of prebiotic chemistry experiments will require standardization to codify the selectivity input by an experimenter (e.g., by parameterizing reaction process conditions). An opportunity for this exists in the field of digital chemistry, where the implementation of a universal programming language for chemical synthesis[40] could allow standardization of selectivity across different experimental protocols, allowing assessing the relative productivity of different prebiotic syntheses even in the face of human or robotic intervention.

Another challenge arises in the size of the chemical space available on planets[41]. The reactants and conditions used in laboratory chemistry, including prebiotic syntheses, often have few or no parallels to geochemical processes[42]. Even for low complexity, low molecular weight molecules we might expect to be geochemically produced, the number of possible molecules far exceeds what can be readily fingerprinted by chemists in the lab[43]. Geochemists are used to this mess, and the challenges of characterizing chemistries in natural environments, but laboratory chemistry often starts from very restricted conditions that are a far cry from this combinatorial chaos simply for reasons of practically. Cheminformaticians are probably the most intimately aware of how vast the space of chemistry really is, studying what they call "chemical space", defined as the space of all possible molecules. It is not possible to estimate the full size of chemical space, but we can estimate the size of subregions. For example, an estimated $10^{60}$ molecules are possible, composed of just the elements C, O, N, S made of up to 30 atoms[44]. By contrast, there are only ~$10^3$ cataloged small molecules in the Kyoto Encyclopedia of Genes and Genomes (KEGG)[45] database, which catalogs known metabolites. Back of the envelope calculation would put metabolism at occupying approximately $10^{-57}$ of the size of just this small subset of chemical structures (this is a very rough estimate as KEGG molecules include other elements).

One might think of the prebiotic Earth as a combinatorial search engine[41] with unconstrained synthesis of diverse small molecules across different geochemical environments[46]. Meteoritic samples are some of the most pristine examples of abiotic chemistry, and tens of thousands of individual compounds have been identified, with many more below detection limits[43]. The question of the origin of life then reduces to how selection could lead to construction of organized molecular systems out of this vast space, and more specifically, the ones we find in extant life. It is relatively easy to target fingerprinting molecules of life as we know it in extraterrestrial samples[47], and this can place bounds on some aspects of the problem. But the combinatorial space needs to be considered, because post-selecting on extant life's building blocks ignores the bulk of chemical diversity in these samples and does little by way of informing why these molecules should be the ones deemed 'special' by entities that evolved from them, who now attempt to look back at their own origin billions of years hence. It is an open question how much of prebiotic synthesis could happen 'in the wild' in natural geochemical environments[48].

Considering the sheer size of chemical space available on planets raises intriguing questions about how much 'information' selective and evolutionary processes can generate. A stunning example of selection against the odds comes from discovery a functional ribozyme from a vast, random sequence pool[49]. Considering the information reduction necessary to find a functioning molecule out of such a large sequence space of possibilities allows the concepts of information theory to be brought to bear on the fitness of evolutionary outcomes[50]. An example is in the quantification of the functional information content of a RNA molecule that binds GTP[51]. While this kind of information theoretic quantify could be of interest as a philosophical concept, it brings to the forefront the key challenge of any information theoretic approach in advancing the science of origins of life: it is only possible to calculate the full size of the space of structures evolution selects from in very special cases[52], and one must post-select on known function. Given how the size of chemical space cannot be computed, nor can the full space be experimentally explored, assigning probabilities across all molecules in this space is not possible.

Despite this fundamental challenge, many promising avenues of research have aimed to quantify the complexity of evolved structures (e.g., those that life produces) using information theoretic quantities. If one considers that evolution must operate by retaining memory and executing this memory to generate structures that can persist, in turn generating further complexity, the process can be analogized to computation[9]. Early on, Schrodinger pointed out the genetic material must be an 'aperiodic crystal' because such a structure would be algorithmically incompressible and have a high information content[4]. While it is possible to treat a genome as a data string on a computer, and analyze it with algorithmic information[53,54], this ignores nearly all other physical details that make genomes the most complex molecular architectures ever identified. Genomes are first and foremost physical materials, and Schrodinger's appeal to an 'aperiodic crystal' was likely intended to treat their information content as a material attribute, not a computational one. Similar thoughts about the materiality of genomes were echoed by Cairns-Smith in his work hypothesizing minerals were the first genetic material[55], where he argued that the aperiodicity of mineral lattices, with defects, could have provided the earliest form of information propagation necessary to mediate a kind of heredity in chemical systems[56]. While we now understand rather well specific cases of nucleic acid polymers and related chemical structures directly templating their information content, including between closely related chemical structures[57], there is no general theory for information transfer in aperiodic organic or inorganic materials that might inform informational transitions between minerals and primitive organic genes.

Algorithmic complexity[37-38] is defined as the length of the shortest computer program, which when run will produce the given output (assuming a fixed programming language), and it is generally uncomputable. It therefore cannot be approximated in the systematic sense required for scientific data analysis (if it were possible to approximate it, it would not be uncomputable). However, other compressibility measures have been used to study phase transitions[58], and evolutionary processes[59]. Data encoding presents a challenge, and another place where human intervention can become problematic in applying such methods to the origin of life[60], because these methods require predefining the relevant state space, or bounding the experimental regime of interest, a challenge for prebiotic experiments. Thermodynamic depth[61] was introduced to avert this issue, defining a physical (rather than computational) notion of complexity in a continuous function of the probabilities of the experimentally defined trajectories leading to production of a given configuration, which might be observed in a lab. However, thermodynamic depth depends on what experiments are done (e.g. via empirically observed frequencies). We again encounter the general challenge for any entropic approach to quantifying complexity, because counting the appropriate state space is often based on what can be observed, and we cannot, even in principle, have access to the full space chemical evolution carves trajectories through, it is too big[52,69]. Considering its application to a chemical system, for example, the thermodynamic depth would depend on the probability of the reaction pathways that produced it[66]. This is untenable as a complexity metric for life detection because the same chemistry will have a different complexity in different contexts and therefore is not standardizable across different origin of life experiments, nor different planetary environments we might aim to test hypotheses about the origins and distribution of life in the universe.

This brings us back to the central thesis of this essay: metrology is critically important to advancing origin of life science. The forgoing issues surrounding understanding the emergence of life, and the complexity this process generates, are challenging and multi-faceted. But more recent work is making traction, opening new avenues for measurable ways to probe the complexity of molecules that could allow major advances. There is a long history of attempts to formalize molecular complexity, but these, like computational measures of complexity, have mostly been based on informational and graph-theoretic formalisms[62,63], and are often applied in drug discovery[64]. Assembly theory (AT) was developed to specifically address the problem of how one might measure the emergence of complex molecules from unconstrained chemical systems in the lab[65]. It's central complexity measure, the assembly index, see **Figure 1**, was derived by working backwards from what can be measured in a mass spectrometer, capturing how evolution constructs complex objects in a way that could be measured and therefore tested[9]. This forms the foundation of the first hypothesis tested with the theory, that is, that some molecules are so complex (as quantified in AT) that they are not producible outside of life. This was confirmed with the early experimental tests of AT, which set an empirical bound of $a \gtrsim 15$ for organic chemistry[12], see **Figure 1**.

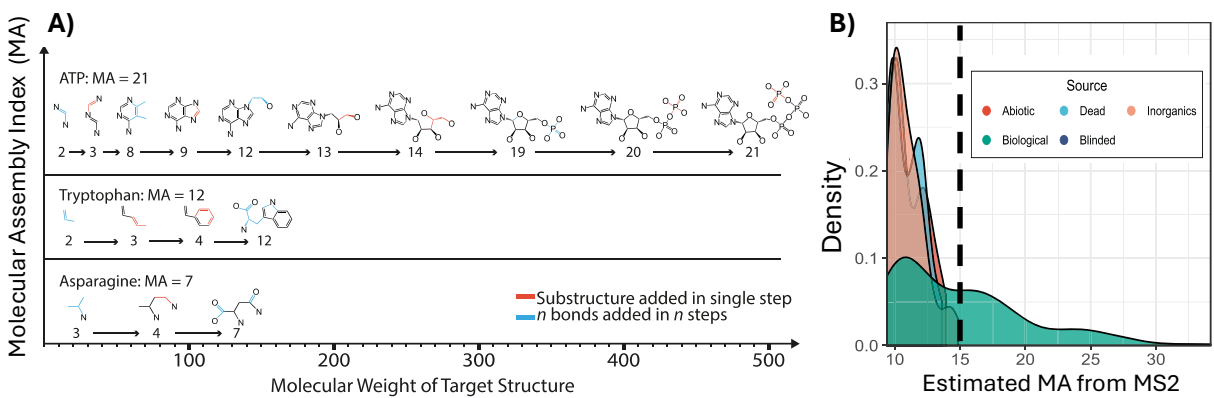

*Figure 1: A) Molecular assembly index (MA) and assembly paths for amino acids asparagine (MA = 7) and tryptophan (MA = 12), and the important energy carrier adenosine triphosphate (MA = 21) with hydrogens and charges omitted for clarity. B) Distributions of MA values estimated from diverse samples in the parent mass range of 300 – 500 m/z determined using mass spectrometry. The biological samples exhibit a wider distribution, showing only biologically produced samples produce MA above a threshold value empirically determined for the aqueous organic chemistry studied of MA ~ 15. Figure adopted from Marshall et al.[66]*

AT's two observables are the assembly index and copy number[67]. Assembly index ($a$) is defined as the minimum number of recursive pairwise operations necessary to construct an object from elementary parts. In molecular assembly theory these parts and operations are chemical bonds[68] (**Figure 1**). Unlike information theoretic measures, the assembly index is built on finite support, without need to assume probabilities over an undefinable space. Measurement of assembly index as an intrinsic property of molecules is validated by several experimental techniques, including mass spectrometry, NMR and infrared, which all yield consistent values for its measurement[69,70]. The other critical observable is copy number. In Sharma et al.[67], the Assembly Equation was introduced, which provides a generalized quantification of the selection necessary to produce an observed configuration of objects:

(1)
$$A = \sum_{i=1}^{N} e^{a_i} \left( \frac{n_i - 1}{N_T} \right)$$

Of note this quantity is exponential in assembly index and linear in copy number, leading to an expected phase transition between abiotic and biotic, **Figure 2**. In experimental detections of molecules, for example in a mass spectrometer, copy number must already be high to be above the threshold for detection. Thus, application of the theory to life detection, as in Marshall et al.[66], focused on direct detection of high assembly index molecules as these would be above the copy number threshold and indicative of high *A*.

The utility of this approach lies in its general applicability. Independent of specific hypotheses about the origin of life (genetics-first or metabolism-first), scientists in the lab can search for, and compare, complexity generated across different experiments and diverse planetary environments in a standardized way. Laboratory implementation measuring molecular assembly with mass spectrometry, NMR and infrared techniques does not require prior knowledge of structure, therefore making the technique applicable to characterizing the complexity of 'messy' chemistries that are hard to precisely fingerprint by comparison to known molecules.

An exciting frontier is expanding the metrology of complexity to other substrates relevant to the origin of life like minerals and polymers. If a common theoretical framework can be identified, grounded in metrology and applicable to quantifying complexity across different material substrates important to the origin of life, we might start to map the diverse major transitions in molecular organization between geochemistry and LUCA using a standardized measurement scheme, and subject to empirical testing.

One frontier is the aforementioned problem of characterizing solid state material complexity, as one might find in minerals. Mineral complexity, like many other forms, has traditionally been studied in the context of information-theoretic approaches[71]. Often the unit cell is used as the relevant mathematical quantity for analysis, but it is not directly observable as an isolated unit of selection in real materials. This led to some debate in the literature[72], pertaining to application of assembly theory to minerals[73]; however, any resolution to the debate should come from what is empirically testable, meaning one must consider the nature of real physical materials, and not solely data encoded in our abstract mathematical representations of them. In this regard, minerals are deeply interesting as a frontier in exploring the trade-offs between copy number and assembly index in **Eq. (1)**. Copy number should be defined by natural correlation length scales within the material that define at what scales selective constraints can produce repeating structure, which with minerals, will occur across multiple scales due to random defects[74]. This presents an exciting frontier because quantifying the selection within minerals would have major implications for understanding their role in the origin of life. For example, if there is an upper bound to the Assembly (**Eq. (1)**) natural materials can encode, due to limitations on the reproducibility of aperiodic defects, this might necessitate an early major transition in molecular structure from a mineral to a covalently assembled genome.

Many different hypotheses for the origin of life have been historically siloed in their approach[26]. Considering the combinatorial complexity of geochemistry and how we can now start to measure selectivity within such large chemical spaces opens new frontiers for unifying currently disparate hypotheses into an evolutionary paradigm for how matter transforms to life. Key insights come from thinking about the origin of life not just as one transition, but as a series of major evolutionary transitions in molecular order[75], where perhaps each has its own properties. For example, each such transition might be characterized as its own phase transition, such that the origin of life should be considered as a cascade of successive phase transitions in molecular selectivity, rather than a singular event[76]. A well-studied example is the phase transition marking the onset of macromolecular homochirality[17]. While both left and right-handed molecular forms are possible, the molecular units composing proteins are exclusively L and the sugars in RNA and DNA are exclusively D. This indicates an early symmetry breaking event between L and D forms, because prebiotic conditions produce racemic mixtures (equal L/D) of most molecules in most circumstances[77]. Here the order parameter is the net chiral asymmetry (e.g., $H = L - D/L + D$), which is 0 in the racemic phase and $\pm 1$ when the symmetry is broken. However, even this example may not be a single-phase transition, but a sequence of nested ones related to molecular complexity. Attaining abundant enough chiral molecules to break L/D symmetry already requires a degree complexity, because chiral centers are much more prevalent in larger, more complex molecules[78].

Chirality provides one example where descriptions embedded in the physicality of molecular structure can allow transcending genetics first and metabolism first approaches. If one were to attempt to detail a cascade of phase transitions leading to LUCA, each mediating more assembled systems then the last, one might start with core carbon metabolism[79]. Many theories for the origin of life begin with one or more of the six central carbon fixation pathways known in modern metabolism[16], which include acetate, pyruvate, oxalocetate, succinate and alpha-ketoglutarate as the standard universal precursors of all biosynthesis[76]. Many, but not all, steps in these metabolic pathways have been demonstrated to be possible in the absence of enzymes[22]. These pathways lead to the synthesis of all 20 coded amino acids and the nucleobases. One set of ideas suggests a layered and scaffolded architecture for the early evolution of metabolism, where multiple nested autocatalytic loops of reactions eventually led to the synthesis of small oligomers (short polymeric sequences). With a self-sustaining system that can produce oligomers, one might next imagine selection operating a new-hierarchical scale of interacting polymeric structures, where associations between nucleotide and amino acids sequences started to confer some advantage to persistence, perhaps due stereochemical associations. From here, the translation system would have been the next major transition in this cascade. While we do not know the precise steps, and the outline above is currently only a conjecture, it is now possible with new theory and measurements to start to work toward how layered constraints and selective mechanisms in each of these set the boundary conditions for a phase transition in selection in the next layer.

To determine whether this is merely an exercise in narrative[80], or empirical reality, requires interfacing theory with experiments. Many emerging methods, based on automated chemical synthesis platforms, are allowing broader exploration of the chemical space of planets to search for the first evolutionary steps in the transition from matter to life[40,81,82]. Exciting frontiers also include searching for general and universal principles, like replicating networks of molecules, that seem to not strictly depend on the molecules of life as we know it[83], and can even be inorganic[84]. Artificial life is one of the most exciting fields of research to develop in recent decades, probing fundamental features of life, but doing so primarily through alternative substrates[85], like software or unconventional evolutionary materials. Simulated approaches are also yielding interesting insights, but relevance to the chemical origins of life should be treated with care. For example, a few studies have demonstrated the spontaneous emergence of replication from simple software instructions[86,87], allowing insight into the question of how easy or hard it is to discover self-reproducing entities capable of further evolution from a 'random' prebiotic system[88]. However, even the simplest software code still includes selective design, so these simulations will suffer from the same challenges of parameterizing the information input by the experimenter as we discussed earlier in this essay.

Finally, another frontier is to consider the implications for searching for life, using a metrology of complexity, on other planets. Currently, state of the art in astrobiology targets biomolecules and metabolic products found in life on Earth, severely constraining our search space against the backdrop of the potentially vast chemistries that could be explored on other planetary bodies[89]. Searching for more general signatures of life, even life as we cannot anticipate it, must be rooted in the science of measurement, leveraging spectroscopy to infer the presence of complexity. This is a critically important frontier for exoplanet science, where unambiguous detection of life will likely not occur until we transition to theory-ladden observation[90], well-motivated by underlying, universal physical principles of life in the universe. Piecing together all these parts will be a challenge, but emerging new measurement schemes, which transcend specific disciplinary hypotheses, will allow researchers to understand the physical transformations during the origin of life more directly, with potential for a truly universal theory for the emergence of life that not only explains our own origins, but provides insights to the possibilities for life in the universe.

# Acknowledgements

The ideas presented in this work were informed by conversations with many excellent colleagues, especially Leroy Cronin, Michael Lachmann, Christopher P. Kempes and Paul C.W. Davies. I would like to thank Schmidt Science and the Sloan Foundation for funding support.